# Siloxane-based $^6$LiF composites for flexible thermal neutron scintillation sensors with high efficiency: effects of $^6$LiF crystals size and dispersion homogeneity

S.M. Carturan[1,2], M. Vesco[1,2], I. Bonesso[1], A. Quaranta[3,4], G. Maggioni[1,2], L. Stevanato[1], E. Zanazzi[3,4], T. Marchi[2], D. Fabris[5], M. Cinausero[2], F. Gramegna[2]

1. Department of Physics and Astronomy, University of Padova, Via Marzolo 8, Padova, Italy
2. Laboratori Nazionali di Legnaro - INFN, Viale dell'Università 2, Legnaro, Italy
3. Department of Industrial Engineering, University of Trento, Via Sommarive 9, Povo, Trento, Italy
4. TIFPA, Trento Institute for Fundamental Physics and Applications, Via Sommarive 14, Povo, Trento, Italy
5. INFN – Section of Padova, Via Marzolo 8, Padova, Italy

**Abstract**

The production of flexible and robust thermal neutron detectors with improved properties as compared to the commercial ZnS:Ag based phosphors is here pursued, exploiting a siloxane binder, whose intrinsic properties as related to the chemical features of the functional groups and to the optical properties are investigated and tailored in correlation with the final performances of the detectors. Two different siloxanes either with pendant phenyl groups or with aliphatic groups have been used, the former being intrinsically fluorescent and with higher polarizability than the latter. Moreover, $^6$LiF crystals have been synthesized by co-precipitation method and the solvent/co-solvent ratio has been changed in order to tune the crystal size. Then, the size effect on the detector efficiency to thermal neutrons has been investigated as related to the energy loss of thermal neutron reaction products inside the crystal and the dispersion homogeneity of the crystals into the composite. To complete the characterization of the produced flexible detectors, the response to γ-rays has been measured and compared to a commercial detector. The careful choice of both the base resin and the $^6$LiF crystals size allows to produce flexible detector for thermal neutrons with performances comparable to the commercial standard and with higher mechanical robustness and stability.

## 1. Introduction

The detection of thermal neutrons is increasingly important for several fields of interest, spanning from nuclear waste repositories monitoring, fissile materials illicit traffic inspection at borders, nuclear physics research dedicated plants, with special reference to Radioactive Ion Beams production, Neutron Spallation Source and other safety control related fields [1,2]. On the other hand, very recently it has been proposed that the thermal neutron detection can be exploited to observe and study the plant root system, including root growth and water uptake, through Thermal Neutron Radiography (TNR) [3,4]. Still another field of public interest where neutron attenuation can be profitably used is the water content estimation of river embankments, thus providing a sort of hydrogeological sensor system to prevent landslip and flooding events. Therefore, the possibility to reveal thermal neutrons with good efficiency using low-cost and durable scintillators is highly desirable and stimulates the search for alternatives to the classic $^3$He tubes, which are exceedingly expensive, though highly efficient. From a commercial point of view, a valid alternative is constituted by thin plates of ZnS activated with Ag (ZnS:Ag) mixed with $^6$LiF powder and several sensors based on this concept are now available. Largely used detectors in form of thin plates are EJ-426 (Eljen Technology, Texas, USA) and BC-704 (Saint Gobain, France). The $^6$Li nucleus is sensitive to thermal neutrons through the following capture reaction

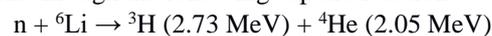
$$n + {}^6Li \rightarrow {}^3H \ (2.73 \text{ MeV}) + {}^4He \ (2.05 \text{ MeV})$$

The generated particles can travel a maximum distance of about 30 µm, triton, and 6 µm, alpha particle, inside the $^6$LiF grain, as calculated by SRIM [5], before losing their whole energy and being stopped. After leaving the $^6$LiF crystals, the generated particles impinge on the ZnS:Ag powder, giving rise to scintillation light emission. Therefore, the size of the $^6$LiF crystals is a critical parameter for the detector performance: the smaller the crystal size, the lower the energy lost, the higher the energy transferable as excitation to the scintillator and, in turn, the higher the final light output. The ZnS:Ag and $^6$LiF mixed powders are dispersed within a plastic binder, acting barely as a support. The dispersion homogeneity of $^6$LiF crystals within phosphor grains is another crucial parameter to minimize light scattering and re-absorption phenomena and, in turn, loss of light output. Finally, the layer must be very thin, in order to minimize γ-rays sensitivity and light loss induced by opacity, whereas loading the binder with as much as possible amount of powder is needed to get maximum detection efficiency, which is around 30% for thermal neutrons on EJ-426 [6]. In the case of self-supporting layers, the detector appears as an off-white, opaque composite, rigid and fragile, in case of self-supporting layers. Most frequently, it is convenient to use the layer deposited on a plastic transparent support, which acts also as a light guide, to be coupled with the photomultiplier tube window. However, the commercial thermal neutron sensors are not easy to handle and care must be adopted in the manipulation of these screens, in order to avoid the development of cracks upon bending and/or under exposure to vibrations. Very recently, the commercial disk EJ-420 as thermal neutron sensor has been coupled to the top surface of a 2" cylinder of scintillator (EJ-299-33A) as γ-rays and fast neutrons detector and this set-up allowed to achieve triple pulse shape discrimination, on the basis of different decay times of the scintillation pulse [7]. Full wrapping of the 2" size cylindrical scintillator EJ-299-33A with EJ-426 thin foils proved to be possible in order to obtain enhanced triple discrimination, but this approach is very challenging in practice, owing to the foils fragility [8]. Another critical issue for commercial thermal neutron detectors is their low heat resistance, which prevents their use in environments where the temperature is higher than in normal conditions, as for example in applications related to oil well logging [9].

The possibility to produce flexible, thermally resistant and easy to handle detectors has been herein explored, by synthesizing $^6$LiF crystals by co-precipitation method [10] and mixing them with ZnS:Ag commercial powder. The preparation of $^6$LiF nanocrystal through different approaches has been previously attempted to obtain transparent composites, in order to exploit the good light output and radiation hardness of siloxane based scintillators doped with suitable fluorophores [11-13]. However, the lack of solubility of $^6$Li complexes led to severe loss of light output.

In this work, a different preparation method has been followed, by mixing $^6$LiF with a high scintillation yield phosphor, namely ZnS:Ag powder, in tenths of mm thin layer of room temperature vulcanizing (RTV) siloxane. The size of $^6$LiF crystals has been changed from few tens of nm up to several μm in order to estimate size effects on the final performances. In addition, two base siloxane matrices either with methyl substituents only or with pendant phenyl groups have been used to investigate both energy transfer effects in case of fluorescent polymer and the role of the polymer structure in the dispersion homogeneity of the inorganic components, $^6$LiF crystals and ZnS:Ag powders.

## 2. Materials and Methods

*2.1 Materials*

$^6$Li metal (enrichment 95%) was purchased from Spectra 2000, while ZnS:Ag (EJ-600) was acquired from Eljen Technology. The resins vinyl terminated (22%-25%) diphenylsiloxane-dimethylsiloxane (PDV-2331 labelled as PSS-22) and vinyl terminated polydimethylsiloxane (DMS-V21) were purchased from Gelest Inc.

*2.2 Experimental*

The preparation of $^6$LiF crystals started from $^6$Li metal that was cut from an ingot and made to react with heated diluted hydrochloric acid, producing $^6$LiCl. To obtain different crystallites size, the produced $^6$LiCl white salt was dissolved in different ratios of water: ethanol mixture, namely 1:0, 1:1, 1:3, 0:1. A solution of ammonium fluoride $NH_4F$ 0.2 M, dissolved in the same mixture, was added dropwise to the $^6$LiCl solution. After few hours of stirring, the solution was centrifuged and the solids washed several times with distilled water; the $^6$LiF crystals were recovered after drying at 40°C under vacuum for several hours.

$^6$LiF powder was mixed with EJ-600 in the weight ratio of 1:3 in an agate mortar to improve homogeneity. The content of solids in each composite was fixed at 70 % wt. with the remaining 30 % wt. composed by viscous resin. Two vinyl terminated siloxane resins were used, with different functional groups in the repeating unit as evidenced in Fig. 1. Two cross-linkers, both based on Si-H reactive groups, were chosen according to the base resin structure, in order to avoid phase separation, whereas the same Pt-based catalyst was used for RTV reaction. To facilitate mixing and casting in the mould, in the case of the highly viscous PSS-22 precursor, some drops of acetone were added to the slurry of powders and resin.

Self-supporting flexible disks of 2" diameter and thickness ranging from 280 up to 350 μm were produced, as can be seen in Fig. 1c, where the photo of a representative sample is reported.

The morphology of the synthesized $^6$LiF powders and the cross-section of the obtained composites were investigated by Scanning Electron Microscopy (SEM-EDS, Tescan) equipped with a detector for backscattered electrons (BSE) and EDS probe (EDAX) for elemental analyses.

Excitation and emission fluorescence of the two series of samples based on either PSS-22 or DMS-V21 were recorded by Jasco FP-6300 spectrofluorimeter, equipped with a 150 W Xe lamp.

In order to estimate the response of the composites to thermal neutrons, each disk was irradiated with thermal neutrons from a $^{252}$Cf source, shielded with 6 cm of polyethylene bricks and placed at 30 cm distance from the front face of the detector. Every disk was coupled to the same Hamamatsu H1949-51R photomultiplier (PMT), using optical grease. To limit external loss of the scintillation light, each disk was covered with an aluminized Mylar sheet and enclosed within a black plastic case, to prevent the interference from external light. A voltage of -1500 V has been applied to the PMT, using a CAEN V6553 bias supply. The anode signal from PMT was read directly by a digitizer, model CAEN V1730, featuring 14 bit resolution with a 2V peak-to peak dynamical range and a sampling rate of 500 MSample/s. Acquisition and signal processing were carried out by a custom software developed at the University of Padova . With the same experimental setup, the γ-rays response of each disk was studied placing $^{137}$Cs γ-rays source at 10 cm distance from the front face of the detector.

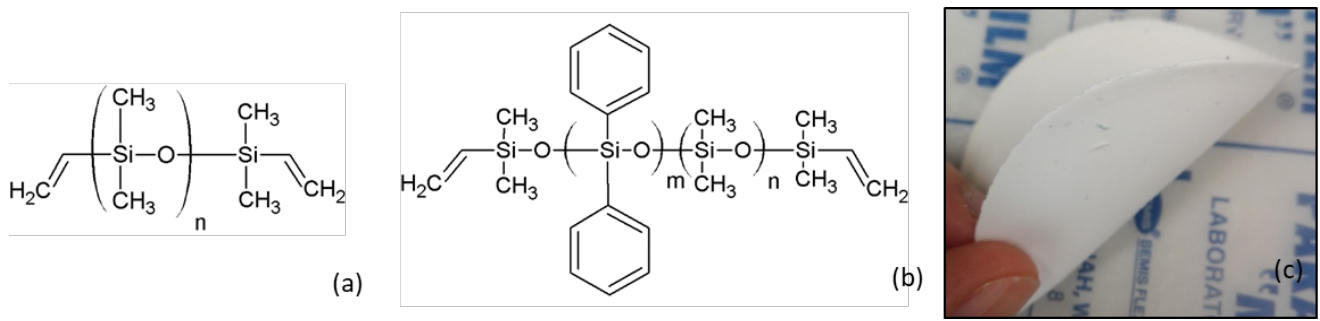

Fig. 1. Chemical structures of the used polysiloxane base resins DMS-V21 (a), PSS-22 (b) and photo of a flexible composite disk (c).

## 3. Results and discussion

*3.1 Scanning Electron Microscopy analysis*

Fig. 2 shows the SEM images of $^6$LiF powders synthesized with different water to ethanol volume ratios. The effect of ethanol as a co-solvent, where $^6$LiF has limited solubility, is clearly evident: in pure water (Fig. 2a) the crystal growth is quite irregular, with formation of acicular structures and incomplete cubic crystals very widely distributed in size, ranging from few µm up to tens of µm. As ethanol is added, the most evident effect is the decrease on average of the crystals size, as can be seen in Fig. 2b and 2c: the average size decreases to less than 10 µm for the 1:1 water:ethanol volume ratio and even to less than 1 µm for the 1:3 ratio. On the other hand, the crystal growth is more regular leading to a narrower distribution in size of regularly shaped cubic crystals. For the synthesis in ethanol only, quite small and regular crystals are produced, with size of few hundreds of nm (Fig. 2d). The precipitation in mixtures of solvents with different polarizability and hydrogen bonding capability is a technique which exploits the variation in solubility of both the reactants and the products for the crystal size control. While LiCl is highly soluble both in water and in ethanol, $NH_4F$ is sparingly soluble in ethanol, therefore the dissolution of $NH_4F$ and formation of solvated F$^-$ ions in ethanol/water solvent mixtures is the critical step in controlling the LiF crystal growth [14]. At the early stages of nucleation, the crystal seeds are surrounded by solvated ions, Li$^+$ and F$^-$, which are attracted toward the core and the crystal grows until the solubility of LiF in that solvent is reached. In pure water, the solubility and the dissociation of the reagents forming solvated ions is the highest, hence crystal growth rate is high and large crystals are formed, with broad size distribution. When ethanol is added, the solubility of $NH_4F$ decreases, and the small core crystal, surrounded by a large amount of non-dissociated molecules, attracts the few F$^-$ ions in the neighbours. As the crystal grows, the concentration of solvated F$^-$ ions decreases, thus promoting further dissociation of $NH_4F$ and reaction progress. In these conditions, the growth is slow and steady, producing crystals with decreasing size as the ethanol fraction increases, until the nm range is reached in case of pure ethanol (Fig. 2d).

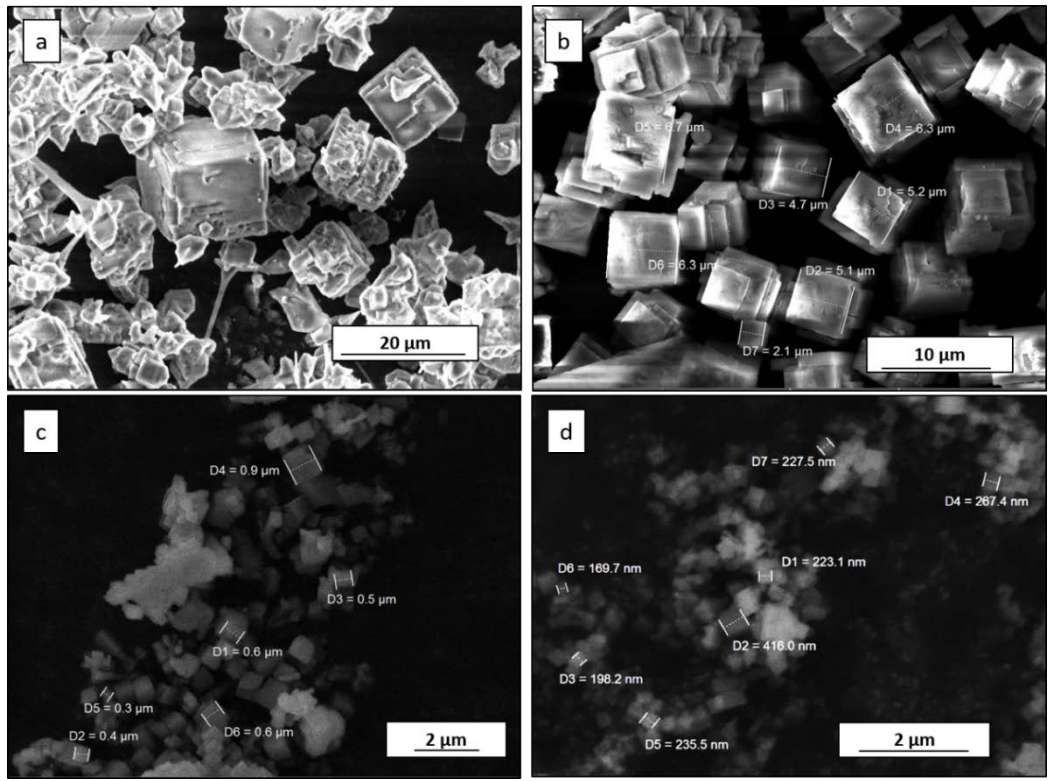

Fig. 2. SEM images of $^6$LiF powders synthesized with different water:ethanol volume ratios: a) 1:0; b) 1:1; c) 1:3; d) 0:1. The dimensions of some crystals are reported in the figures.

Fig. 3 shows the SEM images of the composites cross-section obtained by mixing the $^6$LiF powders, precipitated in different solvents, with ZnS:Ag in weight ratio 3:1, using PSS-22 as a binder (30% wt.). The cross-section of the commercial detector EJ-426 is also shown for comparison (Fig. 3d). The dispersion degree of ZnS:Ag, appearing as roundly shaped micron sized grains in a quite broad range of dimensions, the cubic crystals of $^6$LiF and the siloxane resin can be observed in the different cases. For the composite obtained with crystals precipitated in pure water (Fig. 3a), the large size cubic grains mostly appear as agglomerates, with size in the order of tens of µm, surrounded by resin and phosphor grains. In the case of crystals from precipitated from W:EtOH 1:1 volume ratio (Fig. 3b), an optimal level of dispersion of the inorganic components within the organic binder can be seen. The cubic crystals and the phosphor grains are finely distributed and only few agglomerates larger than some microns are visible. On the other hand, for the sample with crystals some hundreds of nm in size (W:EtOH 0:1 in Fig. 3c), the tiny cubes are highly agglomerated in islands tens of µm wide, and surrounded by the resin, whereas the EJ-600 lumps are visible as clusters of grains several µm far from $^6$LiF agglomerates. A rather unsatisfactory performance of this detector is expected because of the observed microstructure: as said before the charged particles arising from neutron capture events have a range of 30 µm at maximum, hence the energy fraction released into EJ-600 grains, which are several microns far apart, is quite low. Moreover, the aggregation of EJ-600 can cause re-absorption of the emitted light, with a remarkable light output loss. It is worth to observe the microstructure of EJ-426 cross-section in Fig. 3d, where optimal dispersion of the components is present and the crystals of $^6$LiF appear smaller on average than those synthesized by W:EtOH 1:1. Thus, it can be inferred that physical interaction between the ions escaping from the crystal after neutron capture and the scintillator grains is quite effective.

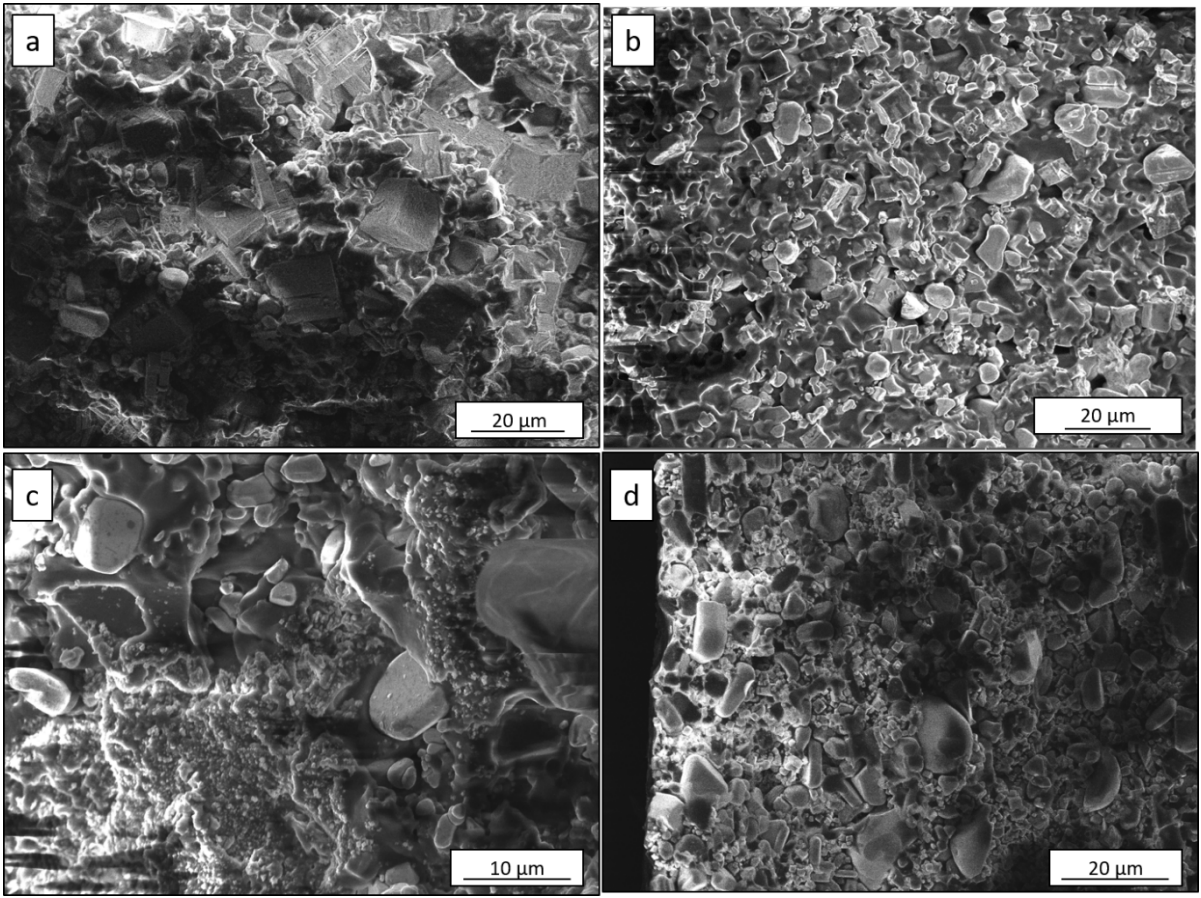

Fig. 3. SEM cross-section images of the PSS-22 composites with $^6$LiF crystal obtained from a) pure water, b) W:EtOH 1:1, c) pure EtOH and d) EJ-426.

In the case of DMS-V21 based composites, the non-homogeneity in dispersion is already visible in the W:EtOH 1:1 samples, as can be observed from BSE images reported in Fig. 4, where DMS-V21 and PSS-22 samples with average micron size $^6$LiF crystals are compared. The detection of backscattered electrons allows to observe the remarkable difference in the dispersion homogeneity in the two different resins, as can be seen from the contrast between $^6$LiF, visible as cubic shaped, dark crystals, and ZnS:Ag, visible as bright grains. It is worth to observe that in PSS-22 the scintillator is finely distributed around the cubic crystals, whereas the use of DMS-V21 causes the formation of aggregates. Several different spots have been investigated by BSE and this scenario has been found almost everywhere along the cross-section of the samples. It can be envisaged that in the case of PSS-22 the close proximity of the cubic grains, emitting ionizing particles, to the luminescent centres and, most importantly, the fine distribution of these components lead to a higher light output on the whole if compared with DMS-V21. In the latter, the aggregation either of $^6$LiF or of ZnS:Ag causes, on one hand, the possibility that the ionizing particles are stopped into the crystals cluster before reaching the scintillator, on the other hand, that light emitted from ZnS:Ag can be re-absorbed, thus leading to light emission loss. The reason for the different morphology of the composites as the base resin is changed is ascribed to the polarizability of the resin itself: in the case of phenyl containing polysiloxane, the dipole-induced ion interaction of the delocalized π electrons with surface ions, either of the LiF crystal or the ZnS:Ag compound, results in the stabilization of the two phases, organic and inorganic, in a way similar to solvation [15]. In case of very small LiF crystals, as those obtained in pure ethanol, the strength of the interaction is not sufficient to prevent agglomeration, thus the phase separation observed in Fig. 3c takes place.

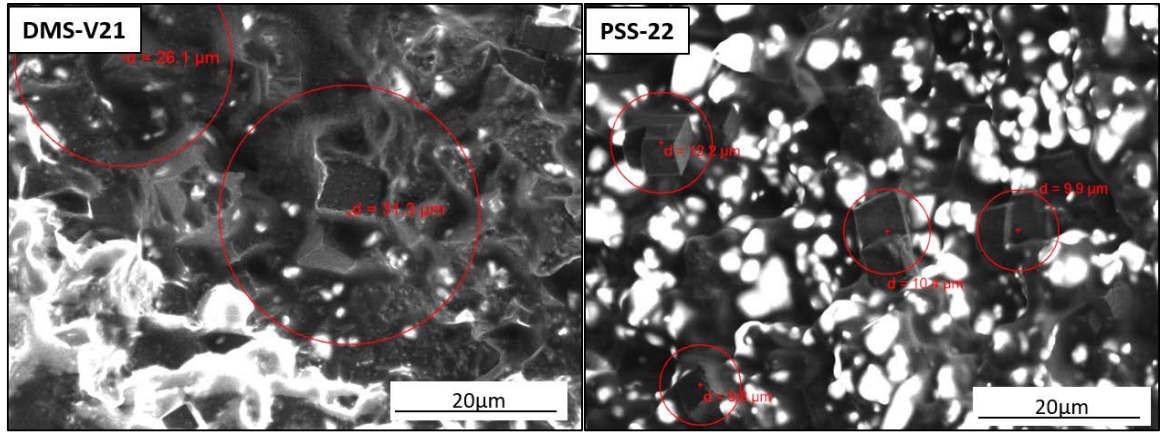

Fig. 4. BSE images of DMS-V21 based (left) and PSS-22 based (right) derived composites cross-section. Both samples contain $^6$LiF crystals synthesized in 1:1 W:EtOH volume ratio.

*3.2 Luminescence analyses*

In order to investigate on the role of the polysiloxane matrix in the enhancement of the scintillation yield a photoluminescence analysis has been performed on both the bare resins and on composites obtained by mixing the resin with 5% wt. of ZnS:Ag. From the emission and excitation spectra reported in Fig. 5, it can be observed that while DMS-V21 shows negligible luminescence, as could be expected on the basis of its chemical structure where chromophore groups are absent, PSS-22 displays intrinsic florescence under excitation at 280 nm, as previously reported in literature [11], with an emission peak centred at around 320 nm. The excitation spectrum shown in Fig. 5 (right) exhibits a broad band, covering a range between 250 and 400 nm with a peak at 340 nm. To this excitation spectrum, corresponds an emission peak centred at 450 nm. Since the excitation region of the phosphor partially overlaps with the emission from the polysiloxane matrix, an energy transfer process promoted by the reaction products crossing the polymer cannot be ruled out. In the reported spectra, the shape and intensity of the EJ-600 emission and excitation features are very similar, irrespectively of the used resin as a binder. The only appreciable difference is related to the excitation band of EJ-600 dispersed in PSS-22, which shows an inflection point in the UV region in correspondence of the excitation spectrum of the binder. This effect can be due to self-absorption effects induced by the phenyl groups of the polymer. Nonetheless, the light emission from ZnS:Ag is so intense that energy transfer effects from the intrinsically fluorescent matrix are not evidenced in this kind of analysis.

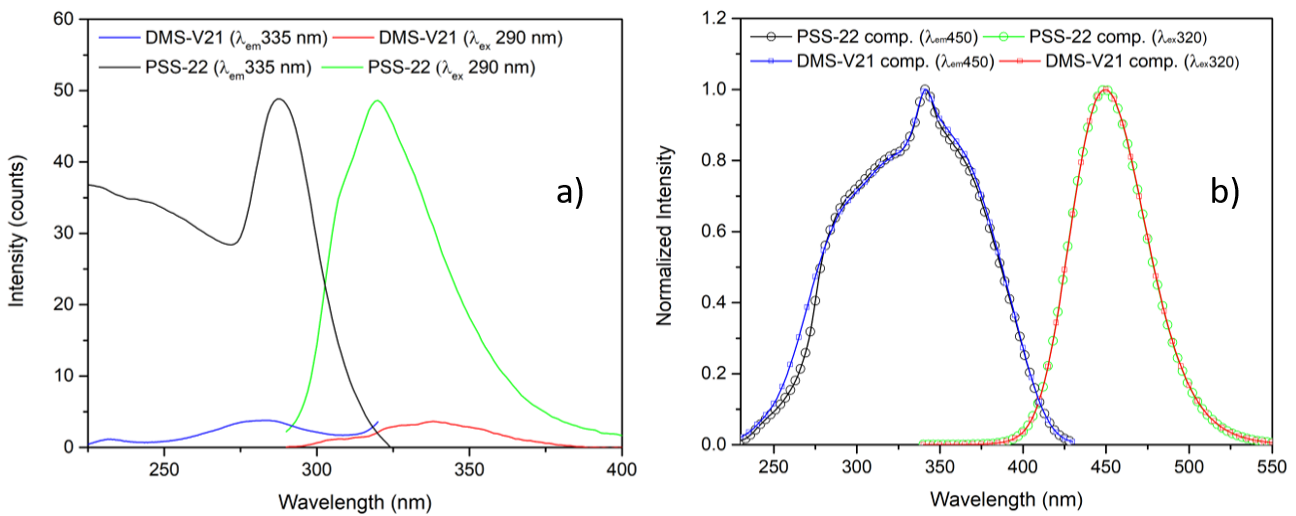

Fig. 5. Excitation and emission spectra collected from the bare resins (a) and EJ-600 containing (5% wt.) composites (b).

*3.3 Thermal neutrons response*

In order to investigate the thermal neutron sensitivity of the produced composites, the scintillation pulses have been analysed through the charge integration method [16]. Thanks to internal FPGA (Field Programmable Gate Array), the digitizer provides online for each event the following key parameters: the time stamp (a reference time from the beginning of the acquisition) and two integrals of the signal (one for the raising part of the signal, $Q_{short}$, and one for the decay part, $Q_{tail}$), as detailed in ref. [17]. From these values, it is possible to define a pulse shape parameter useful to discriminate signals from the nuclei produced through thermal neutron capture reaction from those produced by noise or γ-rays. The Pulse Shape Parameter (PSP) is defined as follows

$$PSP = Q_{tail}/(Q_{short} + Q_{tail}) \quad (1)$$

where $Q_{tail}$ is the integral of the signal tail and $Q_{short}$ is the integral of the fast component. Thermal neutron signals are characterized by higher values of PSP, as defined in Eq. 1, with respect to background noise or γ-rays. In Fig. 6a the time evolution of the pulse intensity related to the detection of thermal neutron from a moderated $^{252}$Cf source in the case of EJ-420 sample is reported. It is worth to observe that the long tail of the falling part of the signal is extending well beyond 1 µs. This particular feature of the pulse shape is related to the excitation density of the ions coming from the neutron capture event, namely triton and alpha. The high local energy deposited by the ions into the ZnS:Ag scintillator produces a high density of excited states, whose interaction slows down the decay to the ground state, giving rise to a longer time than in case of irradiation with photons. In fact, the detection of γ-rays occurs through scattered electrons, producing a much lower excitation density along the path and giving rise to much shorter scintillation decay, in the order of hundreds of ns as previously observed in literature [18,19]. In Fig. 6a the widths of the integration gates for PSP are also reported. In particular, the $Q_{short}$ width was set to 80 ns and the $Q_{long}$ one to 1420 ns. The integration width for $Q_{short}$ starts 40 ns before the overcome of the acquisition threshold, that in the present case was set to 15 mV. In Fig. 6b the plot of PSP versus $Q_{tot}$ is reported for the same sample, exposed to the moderated $^{252}$Cf source for 30 min.

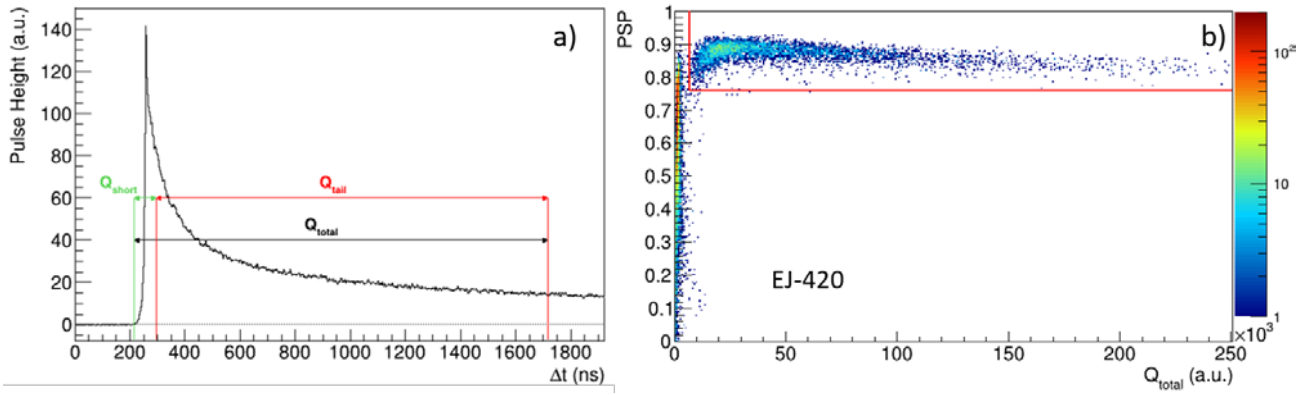

Fig. 6. a) The average signal as pulse intensity versus time collected from the commercial thermal neutron detector EJ-420; b) PSP plot of EJ-420 versus total integration charge ($Q_{total}$) when exposed to $^{252}$Cf source shielded with 6 cm of polyethylene (red lines evidence the region of events ascribed to thermal neutrons).

Thermal neutrons detection efficiency is defined as the ratio between the number of thermal neutron induced events on the detector and the number of impinging thermal neutrons, hence several parameters should be considered, besides the source activity, such as the thickness of the moderator, the distance from the source, the solid angle of the detector and the signal-to-noise ratio that depends on light output, the PMT gain and the acquisition threshold.

In the present experiment, we calculated the relative efficiency of our samples with respect to the commercial standard EJ-420. This relative efficiency has been evaluated by counting the thermal neutrons induced events identified by pulse shape discrimination method in a defined acquisition time interval and in a given detector-source configuration, once the best condition (i.e. acquisition threshold, signal-to-noise rejection, signal integration gates) for signal processing were properly set. The ratio between the thermal neutron counts in different samples with respect to the counts in the standard EJ-420 define our "relative efficiency".

As illustrated in Fig. 6b for the EJ-420 standard, a further software gate in the PSP-$Q_{total}$ plane ($Q_{total} > 7\times10^3$ and $0.76 <$ PSP $< 1.00$) was set for the final selection of thermal neutron induced events and reject others in the background. The 2D plot of PSP-$Q_{total}$ plots for each studied sample are shown in the panels of Fig. 7. The relative efficiency with respect to EJ-420 is reported in Fig. 8a as a function of the W:EtOH volume ratio.

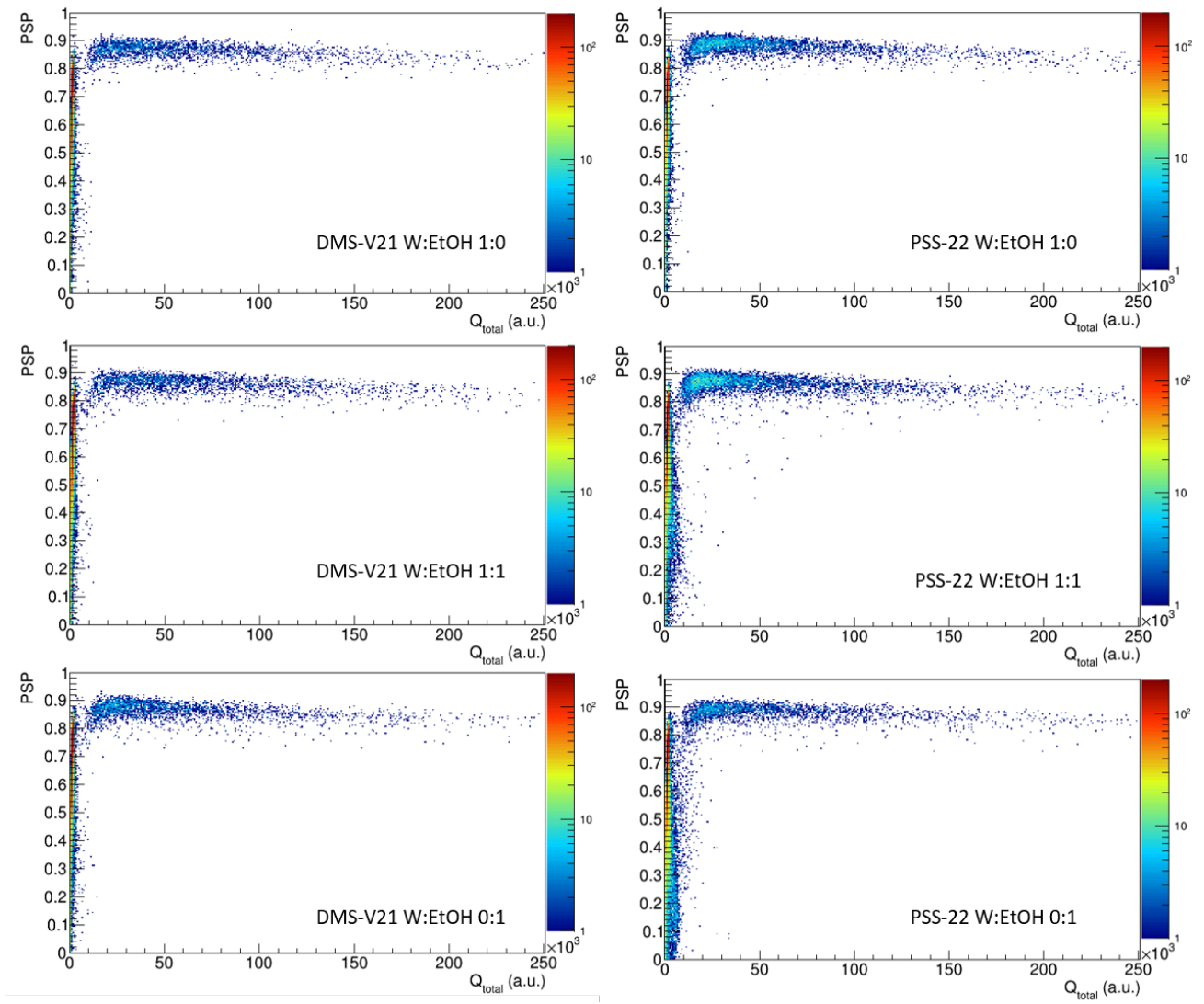

Fig. 7. PSP plots versus $Q_{total}$ of the produced sensors with different average size crystal (decreasing from top to bottom) and different siloxane resins (DMS-V21 on the left column, PSS-22 on the right column).

A remarkable difference in the response between the samples produced with PSS-22 rather than DMS-V21 can be clearly observed. This behaviour was expected from the different distribution evidenced by BSE images, where PSS-22 based composites proved to be more homogeneous in the distribution of $^6$LiF cubic crystals and ZnS:Ag grains. Moreover, a trend can be noticed within the set of PSS-22 samples as regarding the $^6$LiF crystals size: both large size (1:0 W:EtOH synthesis volume ratio) and small size (0:1 W:EtOH synthesis volume ratio) crystals show a lower efficiency with respect to samples containing crystals derived from 1:1 W:EtOH synthesis volume ratio. In the case of smaller crystals, with size of hundreds of nm on average, this is due to the bad dispersion of the powders in the resin as reported in the SEM images of Fig. 3, where the agglomeration of the small cubes in large clusters with size in the order of some microns is clearly observed. In this case, the interaction between the ions produced from capture reactions and the ZnS:Ag grains is definitely less likely to occur. The same detrimental effect occurs in the case of the composite with large crystals, such as $^6$LiF cubes of tens of μm obtained in pure water as solvent system, owing to the long path for the reaction products to exit the crystal and hit the closest ZnS:Ag grain. The best result has been achieved with PSS-22 as base resin, 1:1 W:EtOH synthesis ratio of $^6$LiF, reaching 96% of efficiency with respect to EJ-420, as can be observed in Fig. 8a. The data reported are the average result of five independent measurements obtained by removing the sample, cleaning the PMT glass window and repeating all the procedure to collect the measure, as described previously.

In order to test for the reproducibility of the experimental approach in producing the thermal neutron sensor, three different disks have been prepared using the same synthesis route, i.e. keeping constant the mixing ratio ZnS:Ag: $^6$LiF, the resin % volume, cross-linking conditions and final thickness. Then, each disk has been tested under irradiation with thermal neutrons and the average relative efficiency value of 90 ± 6% has been evaluated from the test as reported in Fig. 8b, leading to the conclusion that the preparation procedure displays optimal reproducibility, owing to the great accuracy in controlling each synthesis step.

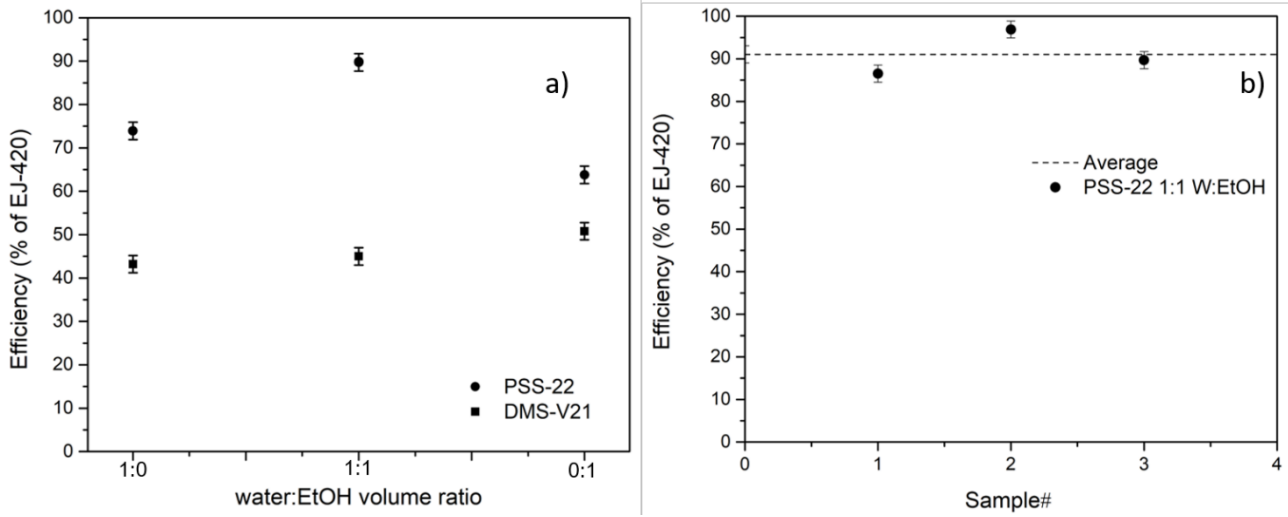

Fig. 8. Thermal neutron detection efficiency relative to EJ-420 of the produced composites either with PSS-22 or DMS-V21 as a function of the volume fraction of ethanol in the synthesis of $^6$LiF crystals (a) and average value of efficiency of three disks prepared in the same way (b). The error bars have been calculated from the dispersions of four independent measurements on the same sample.

*3.4 Sensitivity to γ-rays*

The very low thickness of the prepared detectors should lead to limited sensitivity to impinging γ-rays, which mainly interact with atomic electrons by Compton scattering. Nevertheless, the evaluation of the sensitivity to γ-rays is an important issue to assess the reliability of the detector, since the release of neutrons is often accompanied by γ-rays, leading to spurious signals. In Fig. 9 the pulse height spectra obtained by exposing the standard EJ-420 and the best performing PSS-22 disk to the moderated $^{252}$Cf source are reported. The spectra are the projection on the *x*-axis of the bi-dimensional plots PSP versus $Q_{total}$ shown in Fig. 6b and Fig. 7. The different contributions are separated on the basis of the selection window previously indicated in Fig. 6b, as for both total charge and PSP, hence thermal neutrons are unambiguously identified by their high value of pulse shape parameter [18]. Nonetheless, the possibility that also other interactions, such as background radiation or electronic noise, contribute to the spectrum cannot be ruled out. Therefore, the sensitivity towards γ-rays has been tested by exposing the sample to a $^{137}$Cs source, emitting γ-rays of 0.662 MeV.

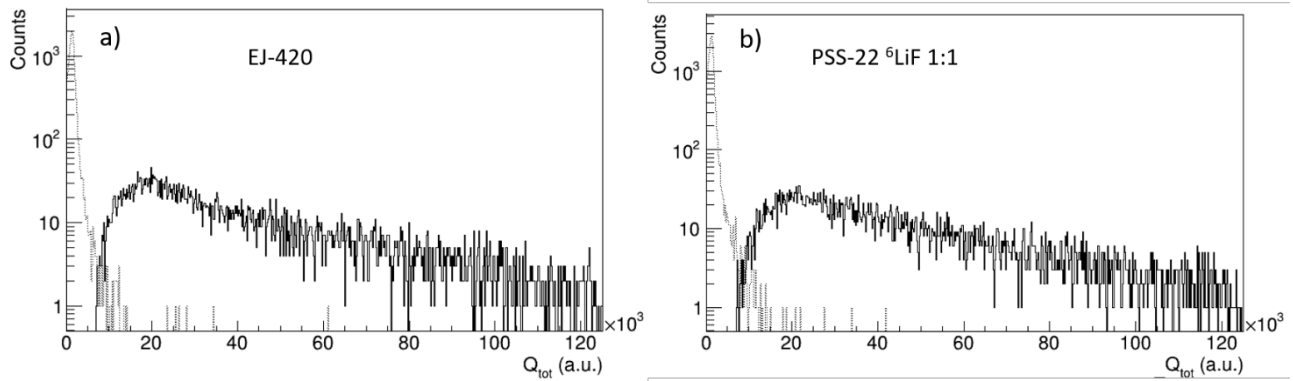

Fig. 9. Pulse height spectra versus $Q_{total}$ showing the different contribution of γ-rays and background radiation (dotted line) and thermal neutrons (continuous line) for a) the standard EJ-420 and b) the best flexible thermal neutron detector herein produced upon irradiation with moderated $^{252}$Cf source.

The same experimental conditions, such as source position, distance of the sample from the source and acquisition time have been kept constant for every disk under test, thus the solid angle has been evaluated as follows

$$\Delta\Omega = 2\pi \left(1 - \frac{d}{\sqrt{(d^2 + R^2)}}\right)$$

with d the distance (10 cm) between the source and the detector surface and R the disk radius (2"). Hence, knowing the source activity and the solid angle ΔΩ, the γ-rays flux from the source was estimated. Then, the sensitivity is defined as the ratio between the total counts recorded by the PMT at a certain threshold voltage for a defined acquisition time and the γ-rays that actually hit the detector.

The sensitivity for EJ-420 and the PSS-22 based disks versus the signal threshold for acquisition is reported in Fig. 10. The most evident result is that the response is remarkably higher for PSS-22 with respect to EJ-420 and for a signal threshold of about 17 mV the difference is one order of magnitude, being $2\times10^{-4}$ for EJ-420 and $2\times10^{-3}$ for PSS-22. This behaviour can be explained on the basis of the chemical composition of the polymers acting as binder of the composites. In fact, since the resin PSS-22 is a silicon based polymer, the Compton scattering cross-section increases with respect to the carbon based epoxy binder used in EJ-420. However, the sensitivity to γ-rays remains very low, thereby highlighting the optimal performance of PSS-22 based composite as thermal neutron detector where interfering signals from γ-rays are negligible.

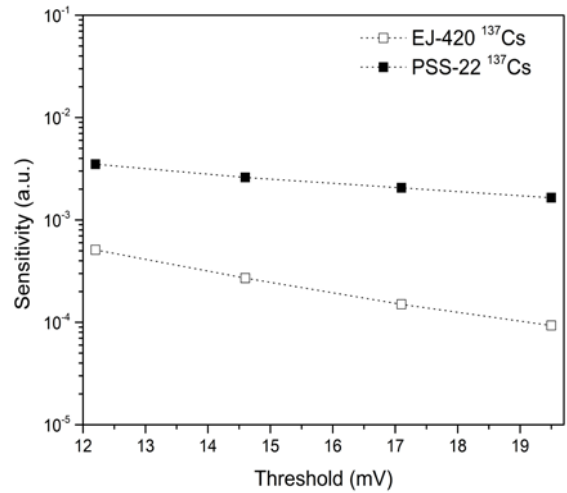

Fig. 10. Sensitivity of the flexible thermal neutron detector based on PSS-22 towards γ-rays from $^{137}$Cs source. The response of the standard EJ-420 in the same conditions is reported for comparison.

## 4. Conclusions

Flexible and robust thermal neutron scintillation sensors have been produced by exploiting the outstanding thermal and mechanical properties featured by silicon based polymers. These detectors showed very similar performances with respect to the commercial ones in terms of neutron counting and low sensitivity to γ-rays. Two polysiloxane resins, namely polydimethylsiloxane and polymethylphenylsiloxane, with different polarizability of substituents have been tested as matrix, entrapping commercial luminescent EJ-600 powder (ZnS activated with Ag) and $^6$LiF crystals as thermal neutron absorber. The crystals have been synthesised by a co-precipitation route and different crystal dimensions ranging from hundreds of nm to tens of μm have been obtained by varying the water to ethanol ratio in the solvent. This synthetic approach has been pursued to evaluate if the decrease in crystals size improves the final light output of the detector, taking into account that the residual energy of the capture reaction products (alpha particles and tritons) escaping from the crystals is strongly dependent on the path through the crystal itself. The results of this work showed that the expected improvement of the light output for smaller crystals can be jeopardized by a bad crystals distribution inside the matrix and the effect is different for the two used resins. In the low polarity polydimethylsiloxane resin, the dispersion of $^6$LiF grains is not homogeneous independently on the used synthetic route, as clearly evidenced by SEM and BSE analyses, and it is characterized by the formation of aggregates whose size is comparable to the range of the reaction products, ultimately portending a remarkable low light output. In the case of the higher polarizability polymethylphenylsiloxane resin, the dispersion is optimal for crystals with size in the order of some μm (W:EtOH 1:1 volume ratio). On the other hand, nanoparticles of $^6$LiF, obtained with a W:EtOH 0:1 volume ratio, undergo clustering and segregation in large agglomerates, thus hampering the exit of ionizing particles and their interaction with EJ-600 grains. BSE analyses have well evidenced these features and the detection efficiency foreseen on the basis of the composite morphology has been confirmed by measurements with a moderated $^{237}$Cf source. Possible energy transfer effects from the fluorescent phenyl containing matrix to the ZnS:Ag luminophore has been also investigated by excitation/fluorescence analyses, but a negligible variation in the emission features has been detected, owing to the very intense luminescence of ZnS:Ag. Therefore, after a careful choice of both the resin and the crystallites size to obtain a good dispersion of the different components, a thin, flexible and mechanically robust detector has been produced, with performances comparable to the best commercially available standard as thermal neutron counter. The γ-rays response has been eventually evaluated, to test for possible misleading signals, and compared to the standard detector. Although a higher sensitivity as related to the standard has been evidenced, the response of the produced detector to γ-rays is still very low and possible interference with the signal from thermal neutrons is negligible. The fabrication of this type of device paves the way towards thermal neutron detection in extremely harsh environment, where resistance to high temperature and intense vibrations is mandatory. Meanwhile, it enables the built-up of multiple radiation detector, where the flexible sensor in the desired shape can be wrapped around or sandwiched between other radiation detectors, albeit maintaining a unique, distinct response in terms of scintillation pulse decay time.

**4. Acknowledgements**

The work presented here has been funded by INFN, III rd Commission (Project NUCL-EX) and by the Department of Physics and Astronomy of the University of Padova (Project BIRD 165833).